\documentclass[osajnl,twocolumn,showpacs,superscriptaddress,10pt]{revtex4-1}

\usepackage{amsmath}
\usepackage{amsfonts}
\usepackage{graphicx}
\usepackage{epstopdf}
\usepackage{hyperref}

\begin{document}

\title{Temporal cloaking with accelerating wave packets}

\author{Ioannis Chremmos}
\email{ioannis.chremmos@mpl.mpg.de}
\affiliation{Max Planck Institute for the Science of Light, D-91058 Erlangen, Germany}


\begin{abstract}
We theoretically propose a temporal cloaking scheme based on accelerating wave packets. A part of a monochromatic lightwave is endowed with a discontinuous nonlinear frequency chirp, so that two opposite accelerating caustics are created in space-time as the different frequency components propagate in the presence of dispersion. The light intensity in the interior of this biconvex time gap is negligible, thus concealing the enclosed events. In contrast to previous temporal cloaking schemes, where light propagates successively through two different media with opposite dispersions, accelerating wave packets open and close the cloaked time window continuously in a single dispersive medium. In addition, biconvex time gaps can be tailored into arbitrary shapes and offer a larger suppression of intensity compared to their rhombic counterparts.
\end{abstract}

\maketitle

Cloaking and invisibility have been particularly popular in recent years. The field gained its impetus from theoretical works that showed how electromagnetic fields can be manipulated with transformation media and made to flow smoothly around objects, thus concealing them from detection \cite{Pendry_2006_EM, Leonhardt_2006_Conformal}. Similar ideas had appeared few years earlier in the context of conductivity problems \cite{Greenleaf_2003_EIT}. Soon after inception, a number of cloaking devices were implemented using electromagnetic metamaterials from microwave \cite{Schurig_2006} to optical frequencies \cite{Cai_2007_Cloak, Valentine_2009, Gabrielli_2009}. A large amount of work has been done since then in the field which has by now reached a high degree of maturity \cite{Zhang_in_Werner_2014}.

Quite recently, the idea of temporal cloaking (TC) emerged with the aim of concealing events rather than objects \cite{McCall_2011}. The concept applies the principles of ray-bending spatial cloaks in space-time: An inhomogeneous time-dependent medium modifies locally the speed of a finite section of a continuous light wave, making its constituent wavelets flow around an avoided space-time region, which stays almost void of optical energy. Nonradiating events taking place inside the medium during this time gap do not interact with the light wave and hence cannot be detected by an observer. The idea lends itself to a multitude of theoretical and practical possibilities ranging from causality editors to handling priorities in digital communications \cite{Kinsler_2014}.

The first demonstration of TC was in a fiber optics setup \cite{Fridman_2012}. Four-wave mixing was used to imprint a discontinuous linear frequency chirp on a finite section of a monochromatic laser signal, thus implementing a time-lens that splits the rays in space-time \cite{Kolner_1989, Bennett_2000_I, Salem_2008}. Transmitting the wave successively through two fibers of opposite dispersion, a time gap opened and closed linearly with distance (rhombic gap), during which the passage of a short pulse was concealed. The chirp was finally removed with a reverse time-lens and the wave assumed its initial uniformity, bearing no evidence of the event or the cloaking process.

The second and latest demonstration of TC applied the same principle to a periodic stream of telecommunication optical pulses \cite{Lukens_2013}. In this work, rhombic time gaps were obtained by means of the fractional Talbot effect and the time-lenses were implemented using electro-optic phase modulators instead of nonlinear parametric processes. Again, in order to close the time gaps, the signal was sent through a second dispersion-compensating fiber.

\begin{figure*}[th]
\centering \includegraphics[width=0.9\textwidth]{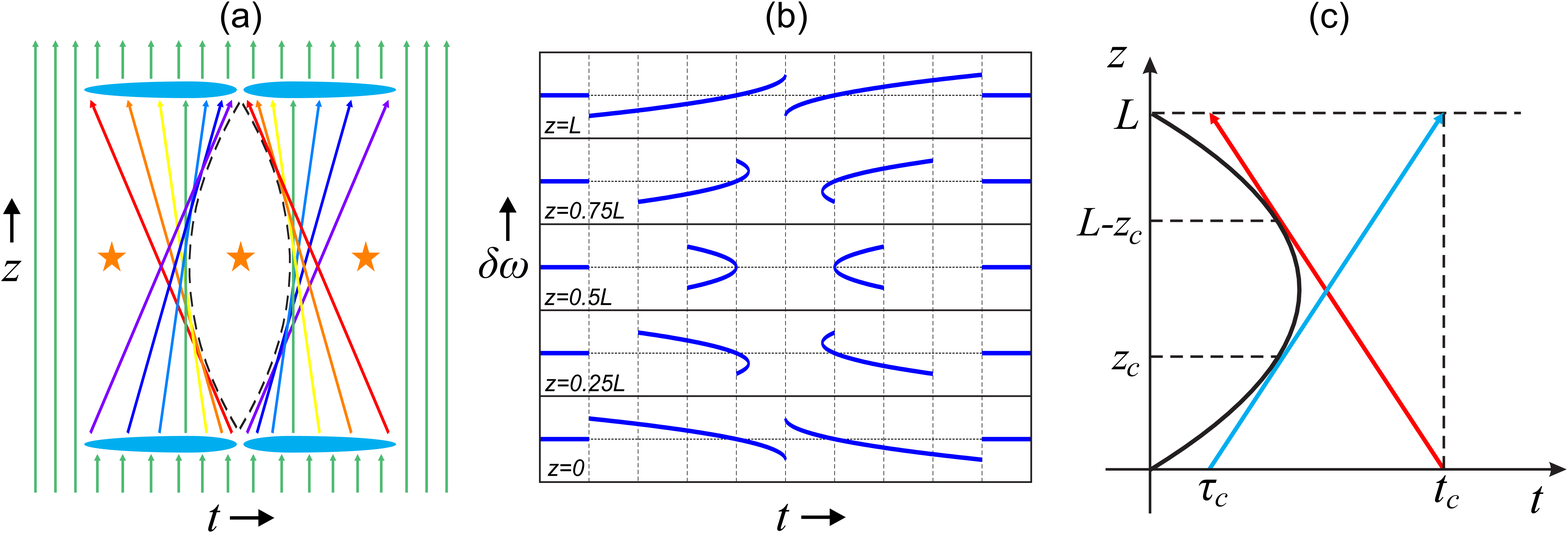}
\caption{(a) Ray schematic of TC with accelerating wave packets. (b) Frequency detuning of the probe wave versus time at different distances. Detunings are measured with reference to the frequency $\omega _0$ of the monochromatic wave, while time is measured in a frame that moves with the group velocity at $\omega _0$. (c) Ray schematic serving the analysis.}
\label{Figure_1}
\end{figure*}

At the heart of TC lies the mathematical equivalence between the diffraction of a paraxial beam in space and the dispersion of a narrowband pulse in a second-order dispersive medium. Further exploiting this analogy, we here propose a TC scheme which derives from the concept of accelerating (curved) optical beams \cite{Siviloglou2007}. We are particularly interested in a special kind of these waves termed abruptly autofocusing (AAF) waves \cite{Efremidis2010}. These have been originally defined as circularly symmetric beams whose amplitude oscillates radially with an Airy profile, so that collapsing parabolic caustic surfaces of revolution are formed as the wave propagates in free space. It has been subsequently shown that AAF beams can be made to successively defocus and focus in an abrupt fashion, thus creating convex ``optical bottles'' which contain negligible energy \cite{Chremmos_2011_Fourier, Chremmos_2012_Autodefocus}. By applying this concept in the spatiotemporal domain, it is possible to design AAF wave packets that create two-dimensional biconvex bottles in space-time in a single medium of positive or negative dispersion. This is different from previous TC schemes, where the cloaked time gap opens and closes in two media of opposite dispersion, including TC in atomic media, where two different pump fields are used to manage the velocities of two detuned pulses \cite{Li_2013_Temp}. Moreover, AAF waves can be tailored into a large diversity of shapes by an appropriate design of their frequency chirp \cite{Chremmos_2011_AAF}. Parabolic caustics, in particular, have the additional advantage of resisting diffraction (or dispersion in time) thanks to their Airy profile.

The proposed scheme is described schematically in Fig. \ref{Figure_1}(a). As in previous configurations, a time-lens is activated for a finite time interval to impart a frequency chirp to a section of a monochromatic light wave (the wave packet). Here, the chirp is nonlinear and particularly designed to make the different frequencies (rays) create two oppositely curved caustics as they travel at different speeds relative to the central frequency $\omega _0$. The evolution of the frequency content with distance is shown in Fig. \ref{Figure_1}(b) in the case of parabolic caustics. The detuned frequencies move away from the center of the wave packet where the chirp is discontinuous, creating a (biconvex) time gap that widens nonlinearly with $z$. At the same time, the detuned frequencies from the rear and front part of the wave packet move toward the center, creating two linearly varying (triangular) time gaps. At $z=L/2$, the three gaps reach a maximum width providing enough time for events to occur without interacting with the wave. The wave packet has now the form of two narrow pulses with two oppositely detuned frequencies arriving at each instant of time (Fig. \ref{Figure_1}(b)). In $z>L/2$, the process is reversed and the time gaps close gradually because of the arrival of rays from the edges of the wave packet. At $z=L$, the gaps vanish and the frequency chirp has the exact opposite of the initial profile. A second reverse time-lens removes this chirp bringing the wave back to its unmodulated form.

To put the above in a mathematical context, we consider the evolution of optical pulses according to the slowly varying envelope approximation \cite{Agrawal_Book}: $2u_z  = i u_{tt}$, where $u(t,z)$ is the pulse envelope, $t_0 t$ is time in a frame moving with the group velocity at $\omega _0$, $z t _0 ^2 / |\beta _2|$ is the propagation distance, $\beta _2$ is the dispersion coefficient at $\omega _0$, while $t_0$ is an arbitrary time scale. Without loss of generality we assume $\beta _2 > 0$, keeping in mind that the same arguments apply equally well in anonalously dispersive media. Being identical to the paraxial equation of diffraction, the pulse equation can be solved by the convolution integral
\begin{equation}
u\left( {t,z} \right) = \frac{1}{{\sqrt {2\pi iz} }}\int\limits_{ - \infty }^{  \infty } {u\left( {\tau ,0^+} \right){\mathop{\rm e}\nolimits} ^{\frac{{i\left( {t - \tau } \right)^2 }}{{2z}}} d\tau }
\label{convolution_integral}
\end{equation}
where $u(t, 0^+)$ is the envelope of the wave packet just after the first time-lens. In studying the propagation dynamics in $0 < z < L$ and in analogy with optical caustics in space \cite{Chremmos_2011_AAF, Chremmos_2011_Fourier, Chremmos_2012_Autodefocus}, it is insightful to use the ray-optics description obtained from a stationary-phase approximation to Eq. \eqref{convolution_integral}. We eventually show that, when the input condition is designed to create two symmetric with respect to $z=L/2$ caustics, the wave evolves to a complex conjugate copy of itself, i.e. $u(t, L^-) = u^*(t, 0^+)$. Subsequently, it can be easily demodulated back to its initial form by passing through a reverse time-lens.

\begin{figure*}[thb]
\centering \includegraphics[width=1.0\textwidth]{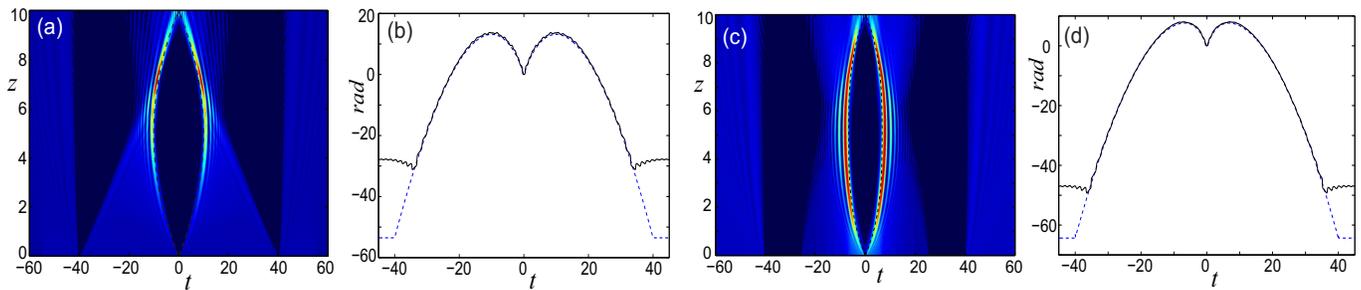}
\caption{(a) TC with a wave packet with input phase $q\left( \tau \right) = 4|\tau| - 8|\tau|^{3/2} /\left( {3\sqrt {10} } \right)$ for $|\tau| \leq 40$, corresponding to the caustic $f(z) = 4 z - 0.4 z^2$ (dashed curves). The input amplitude has been assumed uniform. (b) Phase of the wave packet at the input and output. The solid and dashed lines correspond to $-q(t,L^{-})$ and $q(t,0^{+})$ respectively. (c)-(d) Same information for the caustic $f(z) = 20/3 - (4/75)|z-5|^3$. For the input amplitude we have chosen $a(\tau)=0.5$ in $|\tau>40|$ and $a(\tau)=\sqrt{1-z/10}$ in $|\tau|<40$, where $z$ is obtained by solving numerically $\tau = f(z) - z f'(z)$. The input phase has also been computed numerically.}
\label{Figure_2}
\end{figure*}


We assume that the wave packet exiting the first-time lens is of the form $u\left( {t,0^ + } \right) = a\left( t \right)\exp \left[ {ip\left( t \right)} \right]$, where $p(t)$ is the phase associated with the frequency chirp, while $a(t)$ accounts for amplitude modulation. Substituting into Eq. \eqref{convolution_integral} and using a stationary-phase argument, one finds the equation of rays
\begin{equation}
t = \tau  + q' \left( \tau  \right)z
\label{ray_equation}
\end{equation}
where $q'\left( \tau  \right) = \omega \left( \tau  \right) - \omega _0$ is the instantaneous frequency detuning at time $\tau$. According to Eq. \eqref{ray_equation}, the wave at $(t,z)$ is mainly contributed by a ray starting from $(\tau, 0)$ and traveling at a velocity $(q' (\tau))^{-1}$, relative to the group velocity at $\omega _0$. The starting time $\tau$ is obtained after solving Eq. \eqref{ray_equation} for a given function $q(\tau)$. As understood from Fig. \ref{Figure_1}(a), there can be zero, one or two solutions, depending on the number of rays arriving at $(t,z)$. The regions where no rays arrive have very low optical energy and constitute time gaps for cloaking events. One of the advantages of biconvex over triangular time-gaps is that their inner field decays strongly as an Airy function with the distance from the caustic, a universal property of fold-type caustics \cite{Felsen_Book}.

Rays can be labeled by their starting points $(\tau, 0)$. Alternatively, a ray can be uniquely associated to the point at which it touches the caustic. In present case, there are two mirror-symmetric caustics with respect to $t=0$, so it suffices to study the $t>0$ caustic, assuming a general function $t = f(z)$ that is upward convex $(f''(z) < 0)$. Obviously, the slope of a ray is equal to the slope of the caustic at the touching point, hence the two relations must hold simultaneously
\begin{equation}
\begin{aligned}
 \tau _c  &= f\left( {z_c } \right) - z_c f'\left( {z_c } \right) \\
   q'\left( {\tau _c } \right) &= f'\left( {z_c } \right)
\end{aligned}
\label{tau_c_equation}
\end{equation}
where $(\tau_c, 0)$ is the starting point of the ray which touches the caustic at $(f(z_c), z_c)$ (see Fig. \ref{Figure_1}(c)). The field produced by this ray right before the second time-lens ($z = L^-$) is obtained by a standard stationary-phase integration of Eq. \eqref{convolution_integral}
\begin{equation}
u\left( {t_c ,L^ -  } \right) \approx  - i\sqrt {\frac{{z_c }}{{L - z_c }}} a\left( {\tau _c } \right){\mathop{\rm e}\nolimits} ^{iq\left( {\tau _c } \right) + i\frac{{\left( {t_c  - \tau _c } \right)^2 }}{{2 L}}}
\label{Stationary_phase_field}
\end{equation}
where
\begin{equation}
t_c  =  f\left( {z_c } \right) + \left( {L - z_c } \right)f'\left( {z_c } \right)
\label{t_c_equation}
\end{equation}
is the arriving time of that ray at $z=L^-$. To arrive at Eq. \eqref{Stationary_phase_field}, we have used the relation $q''\left( {\tau _c } \right) =  - z_c^{ - 1}$, which follows by differentiating the second of Eqs. \eqref{tau_c_equation} with respect to $\tau _c$. Note that, being a geometrical optics result, Eq. \eqref{Stationary_phase_field} becomes singular exactly at the caustic $(z_c = L)$. The correct Airy function behaviour of the wave at and through a caustic is retrieved through a higher-order stationary-phase approach \cite{Felsen_Book}.

We now demand that the wave at $z=L^-$ is a complex conjugate copy of the input condition, or $u(t_c, L^-) = u^*(t_c, 0^+)$. According to Eq. \eqref{Stationary_phase_field}, this is equivalent to the conditions: $a\left( {\tau _c } \right)\sqrt {z_c }  = a\left( {t_c } \right)\sqrt {L - z_c }$ for the amplitudes and
\begin{equation}
q\left( {\tau _c } \right) + q\left( {t_c } \right) + \frac{L}{{2 }}\left( {f'\left( {z_c } \right)} \right)^2  = \frac{\pi }{2}
\label{phases_condition}
\end{equation}
for the phases. In the above, we have also used $t_c  - \tau _c  = Lf'\left( {z_c } \right)$ following from Eqs. \eqref{tau_c_equation} and \eqref{t_c_equation}. Differentiating Eq. \eqref{phases_condition} with respect to $z_c$, while computing $\tau _c '(z_c)$ and $t_c '(z_c)$ from Eqs. \eqref{tau_c_equation} and \eqref{t_c_equation}, respectively, we find
\begin{equation}
q' \left( {t_c } \right) =  - f'\left( {z_c } \right)
\label{slope_at_t_c}
\end{equation}
Similarly to the second of Eqs. \eqref{tau_c_equation}, Eq. \eqref{slope_at_t_c} gives the velocity of the ray starting from $(t_c, 0)$. Combining Eqs. \eqref{t_c_equation} and \eqref{slope_at_t_c}, we find that this ray touches the caustic at $f(L-z_c)$ with a slope $-f'(z_c)$. This can happen only if
\begin{equation}
f(z_c) = f(L - z_c)
\label{caustic_symmetry}
\end{equation}
namely if the caustic is symmetric to the line $z= L/2$. For any given function $f(z)$ with this property, the input phase follows by integrating the second of Eqs. \eqref{tau_c_equation} as $q\left( {\tau _c } \right) = \int {f'\left( {z_c } \right)d\tau _c }$, where the function $z_c(\tau _c)$ is obtained by inverting the first of Eqs. \eqref{tau_c_equation}. The result will satisfy Eq. \eqref{phases_condition}, with the exception of some additive constant $q(0)$ that can be absorbed in the input condition. Concerning the condition for the amplitude $a(t)$, it can be shown with the help of Eq. \eqref{caustic_symmetry} that it reduces to $a\left( {\tau _c } \right) = g\left( {z_c } \right)/\sqrt {z_c }$, where $0<z_c<L$ and $g(z_c)$ is an arbitrary function satisfying the symmetry Eq. \eqref{caustic_symmetry}.

It should be noted at this point that, in principle, perfect TC requires time-lenses that can shape both the amplitude and the phase of a signal, which is a demanding task \cite{Weiner_2000}. In practice however, even if the input amplitude is carefully designed, there will always be distortions in the wave arriving at $z=L^-$ due to the discontinuity imparted to the signal by the finite duration of the action of the time-lens. For these reasons, TC works have so far focused on manipulating the spectrum of the wave packet ignoring amplitude variations.

Figure \ref{Figure_2}(a) depicts the evolution of a wave packet whose input phase is designed to create a parabolic biconvex time-gap. As expected, two symmetric Airy-type caustics are formed in excellent agreement with the given analytical parabolas. Figure \ref{Figure_2}(b) compares the phases at $z=0^-$ and $z=L^+$, verifying them as being opposite. In both figures, edge-diffraction effects inevitably distort the edges of the wave packet and the unmodulated signal. To connect with experimental practice, the input amplitude $a(t)$ has been assumed uniform, which brings about some asymmetry in the distribution of intensity with respect to $z=L/2$.

An example with cubic caustics is shown in Figs. \ref{Figure_2}(c,d). With such flatter caustics, the width of the time gap varies slowly around $z=L/2$, which may be useful for cloaking events of the same duration at multiple points along a medium. In this example, the input amplitude has been modulated according to the derived condition $a(\tau) = g(z)/\sqrt{z}$, choosing $g(z) = \sqrt{L z - z^2}$, leading to a symmetric evolution of intensity with respect to $z=L/2$. Note however that, for a given wave packet width, flatter caustics are obtained at the cost of shorter time gaps.

\begin{figure}[t]
\centering \includegraphics[width=0.5\textwidth]{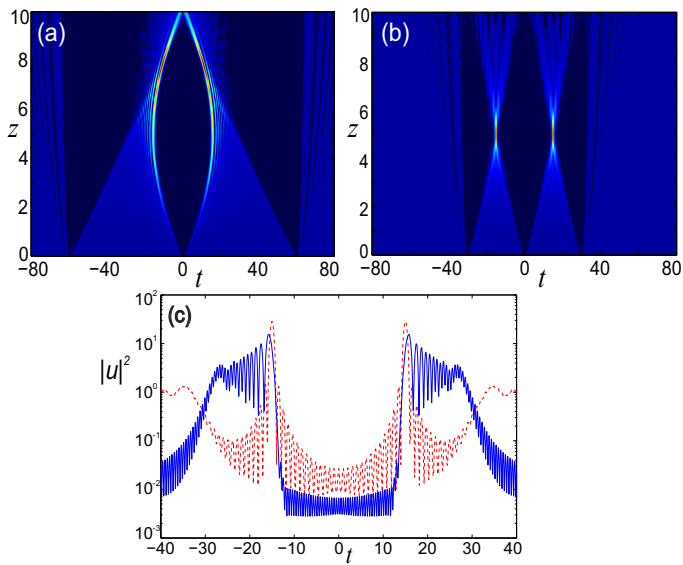}
\caption{Comparison between TC with parabolic biconvex and triangular (non-accelerating) time gaps. (a) Wave packet with input phase $q\left( \tau \right) = 6|\tau| - 4|\tau|^{3/2} /\left( {\sqrt {15} } \right)$ for $|\tau| \leq 60$ corresponding to the parabolic caustic $f(z) = 6 z - 0.6 z^2$. (b) Wave packet with parabolic input phase $q(\tau)= (|\tau| - 15)^2/10$ in $|\tau| < 30$ and constant elsewhere. (c) Intensities of (a)(solid blue) and (b) (dashed red) at $z = L/2$.}
\label{Figure_3}
\end{figure}

We finish by comparing the suppression of optical intensity in the cloaked region between biconvex and rhombic time gaps. The latter are obtained as the zero-acceleration limit of the former, whereby the input chirp becomes linear. Figure \ref{Figure_3}(c,d) compares the intensities at $z=L/2$ between the AAF wave of Fig. \ref{Figure_3}(a) and a zero-acceleration counterpart keeping the propagation length fixed. For a fair comparison, the input phase has been designed to yield the same time gap while the amplitude is uniform in both cases. Clearly, the biconvex gap achieves a better extinction of intensity due to the strong decay of the field through a caustic.

In conclusion, we have theoretically proposed a TC scheme based on accelerating AAF-type wave packets. In contrast to previous TC schemes where rhombic time gaps are created and eliminated in two oppositely dispersive media, accelerating waves create biconvex time gaps continuously in a single medium of positive or negative dispersion. The shape of the gaps can be controlled arbitrarily. More exotic acceleration paterns can be obtained if third order dispersion is also present \cite{Driben_2013}. These ideas count on the implementation of time-lenses that impart a particularly designed nonlinear frequency chirp on optical pulses. Such lenses have indeed been demonstrated in the context of producing accelerating Airy pulses using techniques such as the grating-telescope combination \cite{Chong_2010_Bullets, Abdollahpour_2010}, multi-stage four-wave mixing \cite{Farsi_CLEO_2013} and pulse shaping \cite{Ament_2011}. Given these developments, TC cloaking with accelerating wave packets is certainly within present experimental capabilities.


\end{document}